\documentclass[preprint,5p,number,sort&compress]{elsarticle}

\usepackage[T1]{fontenc}
\usepackage[utf8]{inputenc}
\usepackage[english]{babel}
\usepackage{color}
\usepackage{amsmath}
\usepackage{bm}
\usepackage{graphicx}
\graphicspath{}
\usepackage{enumitem,kantlipsum}
\usepackage[]{hyperref}
\usepackage[normalem]{ulem}

\journal{Chemical Physics}

\begin{document}

\begin{frontmatter}

\title{Tangent space formulation of the
  Multi-Configuration Time-Dependent Hartree equations of motion: The
  projector--splitting algorithm revisited}

\author[guf]{Matteo Bonfanti \corref{cor1}}
\ead{bonfanti@chemie.uni-frankfurt.de}
\author[guf]{Irene Burghardt \corref{cor1}}
\ead{burghardt@chemie.uni-frankfurt.de}

\address[guf]{
Institute of Physical and Theoretical Chemistry, Goethe University Frankfurt,\\
Max-von-Laue-Str. 7, D-60438 Frankfurt/Main, Germany
}
\cortext[cor1]{Corresponding author}

\begin{abstract}
The derivation of the time-dependent variational equations of the
Multi-Configuration Time-Dependent Hartree (MCTDH) method for high-dimensional
quantum propagation is revisited from the perspective of tangent space
projection methods. In this context, we focus on a recently introduced
algorithm [C. Lubich, Appl. Math. Res. eXpress 2015, 311 (2015), B. Kloss et
  al., J. Chem. Phys. 146, 174107 (2017)] for the integration of the MCTDH
equations, which relies on a suitable splitting of the tangent space
projection. The new integrator circumvents the direct inversion of reduced
density matrices that appears in the standard method, by employing an
auxiliary set of non-orthogonal single-particle functions. Here, we
formulate the new algorithm and the underlying alternative form of the
MCTDH equations in conventional chemical physics notation, in a complementary
fashion to the tensor formalism used in the original work. Further, 
key features of the integration scheme are highlighted.
\end{abstract}

\end{frontmatter}



\section{Introduction}
\label{sec:intro}

The Multi-Configuration Time-Dependent Hartree \break (MCTDH) method
 \cite{Meyer1990TheApproach,Manthe1992Wave-PacketNOCl,Beck2000TheWavepackets}
and its multi-layer (ML-MCTDH) variant
\cite{Wang2003MultilayerTheory,Manthe2008ASurfaces,Vendrell2011MultilayerPyrazine}
are powerful methods for quantum propagation in many dimensions. A number of
recent contributions in the mathematical literature have reviewed these
methods from the viewpoint of low-rank tensor approximation techniques
\cite{Koch2010DynamicalApproximation,Lubich2013DynamicalTensors,Bachmayr2016TensorEquations}.
Among these developments, Lubich \cite{Lubich2015TimeDynamics} proposed a
novel MCTDH integration algorithm, which was later implemented and tested on
low-dimensional model systems \cite{Kloss2017ImplementationApproach}. This
algorithm relies on the splitting of the tangent-space
projection \cite{Lubich2015TimeDynamics} and is, hence, termed
projector-splitting integrator. This new integration scheme is
one focus of the present work.

The aforementioned tangent space concept
\cite{Lubich2015TimeDynamics,Koch2010DynamicalApproximation} provides the key connection
between the recent, more mathematical developments and
the conventional derivation of the MCTDH equations of motion.
That is, for a given trial wavefunction, the time-dependent variational
  principle naturally introduces a tangent space which defines the
  best approximation to the time derivative of the wavefunction.
This perspective, which is not usually adopted in the context of MCTDH, is a
useful complement to the conventional derivation of the MCTDH equations of
motion, and provides a natural setting for the derivation of the
projector-splitting integrator. A second focus of the present work is
therefore the introduction to tangent space projections and the unified
derivation of both the conventional MCTDH equations and the modified
projector-splitting version of these equations from this alternative
perspective.

Within the standard MCTDH approach, the equations of motion for the
single--particle functions (SPFs) -- {\em i.e.}, the time-dependent basis of
MCTDH -- {have a highly nonlinear structure, involving a
   time-dependent {\color{black} subspace} projector and the inverse of a
   single-particle density matrix, $\bm{\rho}^{(\kappa)}$. The advantage of
   the projector-splitting integrator lies in the fact that both features are
   circumvented} and the equations are recast in a linear
   form \cite{Lubich2015TimeDynamics} (noting that linearity here refers to the
   form of the equations, while nonlinearity due to the presence of mean-field
   potentials remains a feature of the new scheme). {The price to pay for this
   formal simplification is the introduction of an auxiliary
     set of non-orthogonal SPFs.}
Potential ({near-})singularities of $\bm{\rho}^{(\kappa)}$ are dealt with at
the level of a QR decomposition, {\color{black} whose standard implementation
is able to handle} the case of matrices with large condition
numbers \cite{NoteQR,Golub}.

An extensive number of MCTDH
applications \cite{HeidelbergMCTDH4,Meyer2012StudyingMethod}, spread across all
fields of quantum dynamics, show that the method in its original form is
generally robust and convergeable and that the regularized inversion that is
used for ill--conditioned density matrices rarely affects the quality of the
results. However, the hierarchical ML-MCTDH variant was found to be more
sensitive to initial conditions and to the regularization
parameter \cite{Manthe2015TheRevisited}. Furthermore, numerical analysis has
raised some concerns regarding the convergence to the exact solution in cases
where ill-conditioned density matrices appear \cite{Conte2010AnDynamics}.
Indeed, problems were reported in several cases described in the literature,
{\it e.g.}, related to the fermionic variant of {MCTDH ({\it i.e.},
MCTDH-F) \cite{Kato04,Zanghellini04,Nest05}} where the sensitivity of the
results to the regularization parameter is found to be increased for certain
classes of systems \cite{Hinz2016InstabilitiesHartree-Fock}.

In addition to bringing improvements in these specific cases, the development
of novel integration algorithms may suggest new strategies to avoid
singularities in the general context of variational equations of motion. This
is a problem that, {\it e.g.}, seriously affects methods based on
non-orthogonal basis functions such as the Gaussian-based MCTDH (G-MCTDH)
method \cite{Burghardt1999ApproachesMethod,Burghardt2003MulticonfigurationalBath,Burghardt2008MultimodePyrazine,Romer2013Gaussian-basedTheory}
and its variational multi-configurational Gaussian (vMCG)
variant \cite{Worth2004AWavepackets,Worth2003FullWavepackets,Richings2015QuantumMethod}.

As mentioned above, the purpose of the present paper is twofold. First,
  we aim to provide a bridge between the conventional formulation of MCTDH and
  some of the more mathematically oriented developments, which are usually
  formulated in tensor language and use the concept of the tangent--space
  projection in the treatment of time--dependent variational problems
  \cite{Bachmayr2016TensorEquations}. Hence, the derivation of the tangent
  space projector for MCTDH, which was first presented in
  Ref.\ [\citenum{Koch2010DynamicalApproximation}], is addressed in some
  detail.
Specifically, we will show that the tangent space naturally
    splits into a subspace related to the variation of time-dependent MCTDH
    coefficients and a complementary subspace that is related to the variation
    of the SPFs.
  Second, and against this background, we give a detailed account of the novel
  projector splitting
  algorithm \cite{Lubich2015TimeDynamics,Kloss2017ImplementationApproach}
  including salient features of the {\color{black} integration scheme. As a
  key point, we emphasize that the subspace of the tangent space that is
  associated with the SPF variation is split into two
  components \cite{Lubich2015TimeDynamics}, permitting a new partitioning of
  the equations of motion that formally removes the inversion of the density
  matrix.

The outline of the remainder of this article is as follows. In section
\ref{sec:notation} we briefly explain notational issues, and in section
\ref{sec:tangent-space-projection} we review the notion of tangent-space
projections. In section \ref{sec:projector-splitting-eof}, we discuss the
dynamical equations of MCTDH in the form of the projector-splitting algorithm.
Appendix \ref{sec:matricisation-tensorisation} contains a brief key to translation
between the tensorial and standard notation, and 
Appendix \ref{sec:projector_splitting-eom} provides details of the derivation of the
projector-splitting algorithm. In Appendix~\ref{sec:projector-splitting-integration-scheme},
the integration scheme of the projector-splitting equations is detailed.

\section{Notation}
\label{sec:notation}

We start by giving a brief description of the notation that will be adopted in
this paper. Generally, we will adhere to the standard conventions of the MCTDH
literature \cite{Beck2000TheWavepackets}.

We seek a solution to the time-dependent Schr\"odinger Equation (TDSE)
for a multidimensional state $\Psi$ by approximating the Hilbert space as a tensor product
of $f$ subspaces of low--dimensional SPFs.
The wavefunction is then represented according to the usual MCTDH {\it ansatz},
\begin{equation} \label{eq:MCTDH_Ansatz}
    \Psi (\{ r_\kappa \}, t) = \sum_{j_1 = 1}^{n_1} \sum_{j_2 = 1}^{n_2} \dots \sum_{j_f = 1}^{n_f} 
    \; A_{j_1,j_2, \dots j_f}(t) \; \prod_{\kappa =1}^f \varphi^{(\kappa)}_{j_\kappa} (r_\kappa, t) 
\end{equation}
where $\varphi^{(\kappa)}_{j_\kappa}$ is the $j_\kappa$-th SPF for mode $\kappa$ and
$A_{j_1,j_2, \dots j_f}$ is the tensor of the expansion coefficients.
The SPFs are defined to be orthogonal at all times,
$\langle \varphi^{(\kappa)}_{j_\kappa}(t) \vert \varphi^{(\kappa)}_{j'_\kappa}(t) \rangle
= \delta_{j_\kappa j'_\kappa}$, benefitting from the gauge freedom of
the MCTDH {\em ansatz} Eq.\ (\ref{eq:MCTDH_Ansatz}) \cite{Beck2000TheWavepackets}.
{\color{black} This standard gauge also implies that $\langle
\varphi^{(\kappa)}_{j_\kappa} \vert \dot{\varphi}^{(\kappa)}_{j'_\kappa}
\rangle = 0$. More generally, the gauge can be defined in terms of constraint
operators \cite{Beck2000TheWavepackets}}.

In the tensor formulation that is adopted in the mathematical literature, the MCTDH expansion of Eq.\ (\ref{eq:MCTDH_Ansatz})
is equivalently interpreted as a reduction of the dimensionality of the coefficient tensor.
This is made evident by projecting the expansion Eq.\ (\ref{eq:MCTDH_Ansatz}) on a
time-independent product basis $\{ \chi_{i_1}^{(1)} \dots \chi_{i_f}^{(f)} \}$,
\begin{equation} \label{eq:MCTDH_Ansatz_2}
    \Psi (\{ r_\kappa \}, t) = \sum_{i_1 = 1}^{N_1} \sum_{i_2 = 1}^{N_2} \dots \sum_{i_f = 1}^{N_f} 
    \; Y_{i_1,i_2, \dots i_f}(t) \; \prod_{\kappa =1}^f \chi^{(\kappa)}_{i_\kappa} (r_\kappa) 
\end{equation}
with
\begin{equation} \label{eq:tucker_format}
    Y_{i_1,i_2, \dots i_f}(t) = \sum_{j_1 = 1}^{n_1} \dots \sum_{j_f = 1}^{n_f}
    \; A_{j_1,j_2, \dots j_f}(t) \; \prod_\kappa^f U^{(\kappa)}_{i_\kappa j_\kappa}(t)
\end{equation}
where $U^{(\kappa)}_{i_\kappa j_\kappa} = \langle \chi_{i_\kappa}^{(\kappa)} |
\varphi^{(\kappa)}_{j_\kappa} \rangle $ is the representation matrix on the
primitive grid of the $\kappa$--mode SPFs. From a tensor algebra perspective,
Eq.\ (\ref{eq:tucker_format}) is known as {Tucker decomposition} of the tensor $Y_{i_1,i_2,
  \dots i_f}$ into the {\it core tensor} \cite{Bachmayr2016TensorEquations} $A_{j_1,j_2, \dots j_f}$ and the set
of matrices $U^{(\kappa)}_{i_\kappa j_\kappa}$. As the number of SPFs is
obviously smaller than the size of the primitive basis, the Tucker
decomposition entails a reduction in dimensionality of the original tensor,
taking advantage of its possible sparsity.

Following standard practice \cite{Beck2000TheWavepackets}, we make use of {multi--indices} to cast
Eq.\ (\ref{eq:MCTDH_Ansatz}) in a more compact form (omitting the explicit time and coordinate dependence),
\begin{equation}
      \Psi = \sum_{{J}} \; A_{{J}} \; \Phi_{{J}}
\end{equation}
where ${J}$ represents an $f$-dimensional vector of indices \break $(j_1,j_2,
\dots j_f)$ and $\Phi_{{J}} = \prod_\kappa^f \varphi^{(\kappa)}_{j_\kappa}$ represents a
configuration. Due to the orthonormality of the SPFs, the
configurations are orthonormal as well, $\langle \Phi_{{J}} \vert
\Phi_{{J'}} \rangle = \delta_{{JJ'}}$. 

In our discussion of the MCTDH equations of motion, we will make use of two
additional conventions for multi--indices \cite{Beck2000TheWavepackets}. In
situations where a summation is carried out over all indices {except one}, we
introduce a reduced multi-index,
\begin{equation}
{J}^{\kappa} = (j_1,j_2, \dots j_{\kappa-1},j_{\kappa+1} \dots j_f)
\end{equation}
When it is necessary to label a tensor with a multi--index with the $\kappa$-th entry substituted
with another integer $l$, we write the modified multi--index as
\begin{equation}
 {J}^{\kappa}_{l} = (j_1,j_2, \dots j_{\kappa-1},l, j_{\kappa+1} \dots j_f)
\end{equation}
With these two definitions, we can define {single-hole functions} (SHFs) as
\begin{equation}
 {\Psi}^{(\kappa)}_{l} = \sum_{{J}^{\kappa} } \; A_{{J}^{\kappa}_{l}} \;
 \prod_{n \neq \kappa} \varphi^{(n)}_{j_{n}}
\end{equation}
and the wavefunction Eq.\ (\ref{eq:MCTDH_Ansatz}) can be re-written as a
product of SPFs and SHFs \cite{Beck2000TheWavepackets},
\begin{equation} \label{eq:spf_shf_decomposition}
      \Psi = \sum_{l_\kappa} \varphi_{l_{\kappa}}^{(\kappa)} {\Psi}^{(\kappa)}_{l_{\kappa}}
\end{equation}
which is most convenient when equations are defined within a given $\kappa$th subspace.
In terms of the SHFs, we can further write the $\kappa$--mode single--particle density matrix as
the overlap of SHFs,
\begin{equation} \label{eq:shf_overlap}
      {\rho}^{(\kappa)}_{l_\kappa {l}_\kappa'} 
      = \langle {\Psi}^{(\kappa)}_{{l}_\kappa} |  {\Psi}^{(\kappa)}_{{l}_\kappa'} \rangle
\end{equation}
noting that ${\rho}^{(\kappa)}_{l_\kappa {l}_\kappa'} = \langle
\varphi^{(\kappa)}_{l_\kappa'} \vert \hat{\rho}^{(\kappa)} \vert
\varphi^{(\kappa)}_{{l}_\kappa} \rangle$ 
where $\hat{\rho}^{(\kappa)} = {\rm Tr}_{\kappa'\neq \kappa} \{ \Psi \Psi^\ast \}$ is the reduced density operator in the
$\kappa$th subspace.

\section{Tangent-space projection of the time-dependent Schr\"odinger equation}
\label{sec:tangent-space-projection}

The projector-splitting
scheme \cite{Lubich2015TimeDynamics,Kloss2017ImplementationApproach} is best
understood when the equations of motion are derived in terms of a
{tangent-space} projection of the TDSE. This is equivalent to the use of the
Dirac-Frenkel Variational Principle (DFVP) to derive the MCTDH
equations \cite{Beck2000TheWavepackets}. Here, we state the main results and
refer to Ref.\ [\citenum{Lubich2008FromAnalysis}] for further background
from a mathematical perspective.

\subsection{Tangent space projection}

The conventional formulation of the DFVP states that the 
best approximation to the time evolving wavefunction at a given
time $t$ is obtained as the function 
$\Psi$ which satisfies \cite{Lubich2008FromAnalysis}
\begin{equation}
   \langle \delta \Psi | \dot{\Psi} - \frac{1}{\imath \hbar} H \Psi \rangle = 0
\end{equation}
where $\delta \Psi$ is an allowed variation of the wavefunction which is
compatible with the chosen {\em ansatz} (see, {\em e.g.},
Eq.\ (\ref{eq:MCTDH_Ansatz})). 
In mathematical terms,
if $\mathcal{M}$ is the smooth submanifold of the Hilbert space in which we are seeking an
approximation to the time--dependent state, $\delta \Psi$ is an element of $T_\Psi 
\mathcal{M}$, the tangent space of $\mathcal{M}$ at $\Psi$.

In Ref.\ [\citenum{Broeckhove1988OnPrinciples}], it was shown that the DFVP
is equivalent to the McLachlan Variational Principle
(MLVP) \cite{McLachlan1964AEquation}, provided that the tangent space $T_\Psi
\mathcal{M}$ is a complex linear space. According to the MLVP, the best
solution to the time--dependent problem is obtained when we approximate the
exact derivative with a vector $\dot{\Psi} \in T_\Psi \mathcal{M}$ that has
the minimal distance -- with the metric induced by the scalar product -- from
the exact derivative of the
wavefunction \cite{Kucar1987Time-dependentApproach}, which is given by the TDSE
as $({\imath \hbar})^{-1} H \Psi$. That is,
\begin{equation}
  \delta \| \dot{\Psi} - \frac{1}{\imath \hbar} H \Psi   \|^2 = 0
\end{equation}
This ``geometrical'' condition can be solved by 
introducing tangent--space projectors \cite{Raab2000OnPrinciple,Lubich2004OnDynamics,Lubich2008FromAnalysis}
such that, at any time, the best approximate derivative $\dot{\Psi}$ is constructed as the orthogonal
projection of the full derivative $({\imath \hbar})^{-1} H \Psi$ onto the tangent space $T_\Psi
\mathcal{M}$. Hence \cite{Lubich2008FromAnalysis},
\begin{equation} \label{eq:orthogonal_proj_derivative}
  \dot{\Psi} = \mathcal{P}(\Psi) \frac{1}{\imath \hbar} H \Psi
\end{equation}
where $\mathcal{P}(\Psi)$ is the orthogonal projector onto the tangent space $T_\Psi \mathcal{M}$.
Given that $\dot{\Psi}$ is a vector of  $T_\Psi \mathcal{M}$ by construction, it belongs to the
range of the projector $\mathcal{P}(\Psi)$. Thus Eq.\ (\ref{eq:orthogonal_proj_derivative}) can be
rearranged as a projected TDSE \cite{Lubich2008FromAnalysis}:
\begin{equation}
   \label{projected_TDSE}
   \mathcal{P}(\Psi) \left[ \dot{\Psi}  - \frac{1}{\imath \hbar} H \Psi \right] = 0
\end{equation}
In practice, Eq.\ (\ref{projected_TDSE}) is the most convenient form to derive the equations of motion from the MLVP, once 
an explicit formula for the projector $\mathcal{P}(\Psi)$ is known.

Any set of equations of motion arising from a specific
  formulation of Eq.\ (\ref{projected_TDSE}) will satisfy the variational
  principle and, hence, will conserve norm and energy \cite{Lubich2008FromAnalysis}.

\begin{figure}
   \centering \includegraphics[width=0.95\columnwidth]{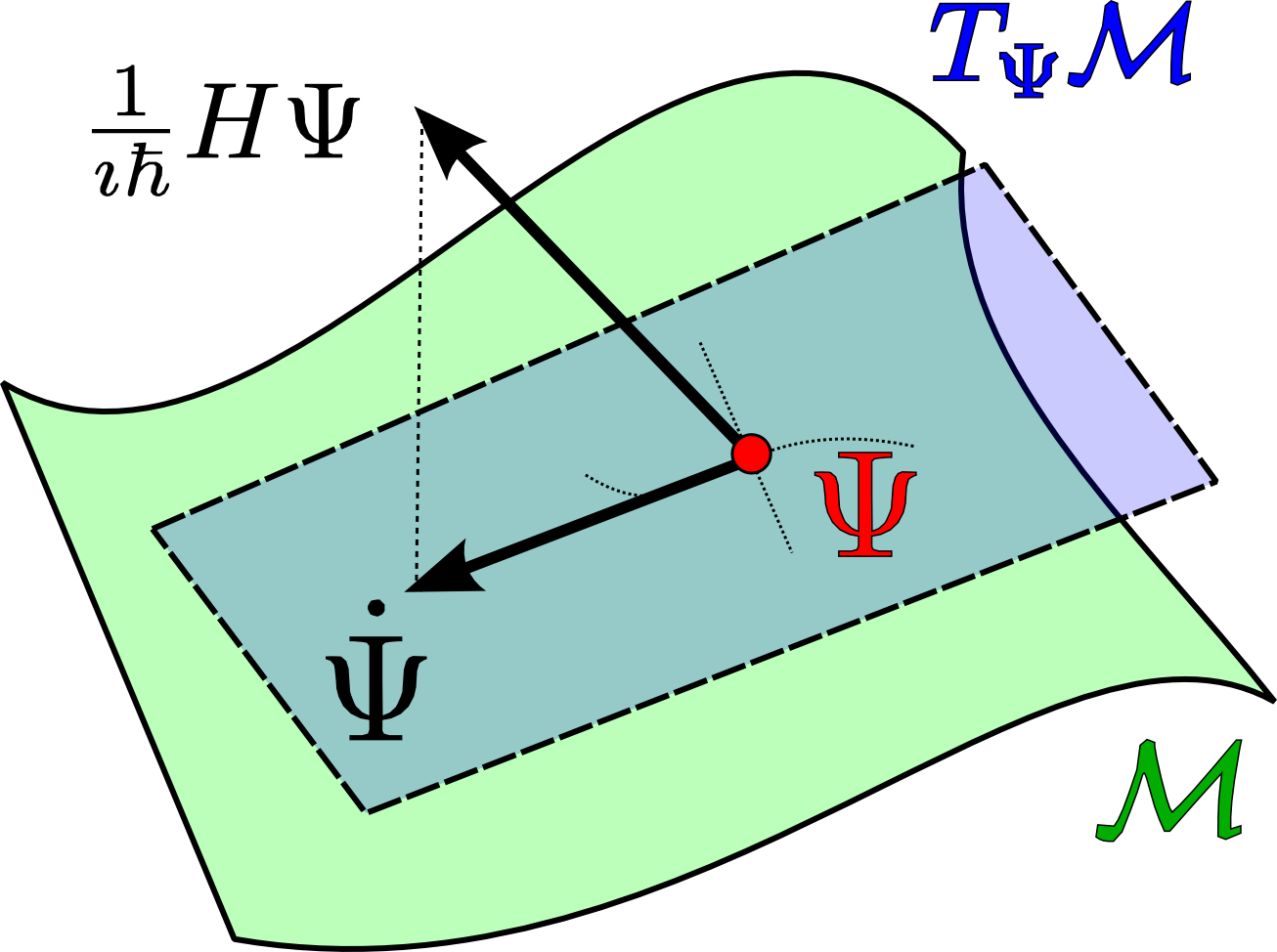}
   \caption{Pictorial representation  of the tangent--space projection. In a given point $\Psi$
            of the variational manifold $\mathcal{M}$ of the full Hilbert space, we construct
            the tangent--space $T_\Psi \mathcal{M}$ as the vector space spanned by the first--order
            variation of the parameters. According to the DFVP, the best local approximation of the
            time--derivative $({\imath \hbar})^{-1} H \Psi$ is given by $\dot{\Psi}$, its
            orthogonal projection onto the tangent space. }
   \label{fig:tangentspaceproj}
\end{figure}

The notion of the tangent--space projection, pictorially represented in Fig.
\ref{fig:tangentspaceproj}, gives an illuminating and immediate understanding
of the variational principle and its implications. When the shape of the
wavefunction is defined according to a chosen {\it ansatz}, we are
constraining the time evolution to a submanifold $\mathcal{M}$ of the full
Hilbert space. The dynamics that is returned by the DFVP is such that at any
given time the derivative of the approximate wavefunction $ \dot{\Psi} $ is
optimal, in the sense that it is closest to the exact value of the derivative
at that point $({\imath \hbar})^{-1} H \Psi$. However, $ \dot{\Psi} $ is
obviously constrained to reside within the tangent space $T_\Psi \mathcal{M}$
since the evolving wavefunction cannot ``escape'' from $\mathcal{M}$. No {\em
global} condition is given for the dynamical propagation; instead, the
approximation is chosen such as to guarantee that at any time the wavefunction
evolution diverges the least possible from the exact dynamics.

{As a consequence, one cannot exclude that small errors that are incurred at
each instant of the propagation may add up to a large deviation of the
dynamics from the exact evolution at longer times.} Of course, as the
submanifold approaches the full size of the space, the projector
$\mathcal{P}(\Psi)$ approaches unity and the approximated evolution tends to
the prediction of the TDSE.

\subsection{Tangent space projection for the MCTDH {\em ansatz}}

From a practical point of view, the construction of the tangent--space
projection for a specific wavefunction {\it ansatz} can be obtained from the
different linearly independent components that constitute the first--order
variation of the wavefunction.

To illustrate the construction of the tangent--space projector, we now turn to
the MCTDH case as an example. In this case, the tangent space is a complex
linear space, since both the $A$ vector and the SPFs are assumed to be
complex-valued. As a consequence, the DFVP and the MPVP are equivalent, and
both are in turn equivalent to a least--action
principle \cite{Broeckhove1988OnPrinciples}.

Referring to Eq.\ (\ref{eq:MCTDH_Ansatz}),
the first-order variation of the MCTDH {\it ansatz} is given by
\begin{equation} \label{eq:MCTDH_first_order_variation}
   \delta \Psi = \sum_{{J}} \; {\delta A}_{{J}} \; \Phi_{{J}} +
   \sum_{\kappa} \biggl( \sum_{l_\kappa}  \; {\delta
   \varphi}_{l_\kappa}^{(\kappa)} {\Psi}_{l_\kappa}^{(\kappa)}
\biggr)
\end{equation}
{\it i.e.}, a sum of terms relating to the variation of the $A$ coefficients
and the SPFs in the $\kappa$th subspaces. As a remark about notation, $\delta$
indicates the variation of the corresponding quantity, specifically ${\delta
A}_{{J}}$ is the variation of the ${J}$--th $A$ vector component, and ${\delta
\varphi}_{l}^{(\kappa)}$ is the variation of the $l$--th SPF for mode
$\kappa$. Eq. (\ref{eq:MCTDH_first_order_variation}) then represents a generic
vector of $T_\Psi \mathcal{M}$. In contrast, the derivative quantities
$\dot{\Psi}$, $\dot{A}_{{J}}$ and $\dot{\varphi}_{l}^{(\kappa)}$ indicate
the specific tangent-space vectors resulting from the DFVP.

The tangent-space projector ${\cal P}$ of
Eq.\ (\ref{eq:orthogonal_proj_derivative}) directly relates to the first order
wavefunction variation of Eq.\ (\ref{eq:MCTDH_first_order_variation}), such
that ${\cal P}(\Psi) \delta \Psi = \delta \Psi$, i.e., the first-order
variation lies in the tangent space $T_\Psi \mathcal{M}$ by construction.
Hence, we naturally aim to construct ${\cal P}$ with a similar partitioning as
Eq.\ (\ref{eq:MCTDH_first_order_variation}) \cite{Koch2010DynamicalApproximation},
\begin{equation} \label{eq:projector_standard}
   \mathcal{P}(\Psi) = \mathcal{P}_0(\Psi) + \sum_{\kappa} \mathcal{P}_{\kappa}(\Psi)
\end{equation}
While it is tempting to construct a one-to-one correspondence between the
components of ${\cal P}(\Psi)$ in Eq.\ (\ref{eq:projector_standard}) and the components of $\delta \Psi$ in
Eq.\ (\ref{eq:MCTDH_first_order_variation}), we will need to make sure that $\mathcal{P}_0(\Psi)$ and
$\mathcal{P}_{\kappa}(\Psi)$ refer to orthogonal projections. In the
following, we use the notation $\delta \Psi = \delta \Psi_0 +
\sum_\kappa \delta \Psi_\kappa$ for the parts of the linear variation that
relate to the subprojections of Eq.\ (15), and we will show below how these
connect to the r.h.s.\ of Eq.\ (\ref{eq:MCTDH_first_order_variation}).

To start with, $\mathcal{P}_0(\Psi)$ is chosen as the projector onto the
configurations $\Phi_{{J}}$,
\begin{equation}\label{eq:p0_projector}
   \mathcal{P}_0(\Psi) = \sum_{{J}} | \Phi_{{J}} \rangle \langle \Phi_{{J}} |
\end{equation}
which, when acting on a generic state, expands this state as a linear combination of 
configurations $\Phi_{{J}}$. The associated portion of the linear variation
Eq.\ (\ref{eq:MCTDH_first_order_variation}), 
here denoted $\delta \Psi_0$, reads
\begin{eqnarray}\label{eq:deltapsi0}
  \delta \Psi_0 & = &  \mathcal{P}_0(\Psi) \delta \Psi \nonumber \\
  \mbox{} & = & 
  \sum_{{J}} \; {\delta A}_{{J}} \; \Phi_{{J}}
  + \mathcal{P}_0(\Psi) \sum_{\kappa, l_\kappa} \;
  {\delta \varphi}_{l_\kappa}^{(\kappa)} {\Psi}_{l_\kappa}^{(\kappa)}
\end{eqnarray}
Since $\mathcal{P}_0(\Psi)$ not only relates to the $A$ coefficient variation
but also acts on the second term corresponding to SPF variations, the
definition of the remaining subspace projections $\mathcal{P}_\kappa(\Psi)$ of
Eq.\ (\ref{eq:projector_standard})
  needs to be chosen such as to
project $\sum_{\kappa, l}  {\delta
  \varphi}_{l}^{(\kappa)}{\Psi}_{l}^{(\kappa)}$ onto
the subspace complementary to the range of $\mathcal{P}_0(\Psi)$:
\begin{eqnarray}
   \sum_\kappa \delta \Psi_{\kappa} & = &
   \sum_{\kappa} {\cal P}_\kappa (\Psi) \delta \Psi \nonumber \\
   \mbox{} & = & 
   \left( 1 - \mathcal{P}_0(\Psi) \right) \sum_{\kappa, l}  \;
   {\delta \varphi}_{l}^{(\kappa)}  {\Psi}_{l}^{(\kappa)}
\end{eqnarray}
With a few lines of algebra, using again the orthogonality of the SPFs, we find that
\begin{eqnarray} \label{eq:variation_second_part}
   \sum_\kappa \delta \Psi_{\kappa}  & = & \sum_{\kappa, l}
   \Psi_{l}^{(\kappa)} \left( {\color{black} {1}} - P^{(\kappa)} \right) {\delta 
\varphi}_{l}^{(\kappa)}
\end{eqnarray}
where 
\begin{eqnarray}
   P^{(\kappa)}= \sum_i | \varphi^{(\kappa)}_{i} \rangle \langle \varphi^{(\kappa)}_{i}|
\end{eqnarray}
is the projector onto the space spanned by the $\kappa$-mode SPFs \cite{Beck2000TheWavepackets}. This is the projector appearing in the
conventional MCTDH equations. 
To underline the difference between the different types of projectors, we
indicate the tangent--space projector and its components with the calligraphic letter $\mathcal{P}$ while the subspace projector is given in Roman type $P$.

Eq.\ (\ref{eq:variation_second_part}) shows that the $\delta \Psi_\kappa$ part of the
first--order variation of the wavefunction is spanned by products between the
SHFs and SPF variations, with the latter being constrained to the orthogonal
complement of the SPFs. In the final equations of motion, this condition will
guarantee that the SPF propagation does not involve changes which are already
represented by the time evolution of the $A$ coefficients. From
Eq.\ (\ref{eq:variation_second_part}) we further infer that variations
corresponding to different modes $\kappa \neq \kappa^\prime$ are orthogonal,
namely
\begin{eqnarray}
\langle \delta \Psi_\kappa \vert \delta \Psi_{\kappa'} \rangle = 0
\end{eqnarray}
as can be seen, {\it e.g.}, by letting the projector $( 1 - P^{(\kappa)} )$ act on the ket ${\Psi}_{l'}^{(\kappa')}$.

From the above, we now identify the projectors
$ \mathcal{P}_{\kappa}(\Psi) $ as
the tensor product of two subspace projectors,
{\color{black} 
\begin{eqnarray}\label{eq:pkappa_projector}
  \mathcal{P}_{\kappa}(\Psi) = \left( {\color{black} {1}} -  P^{(\kappa)} \right) \otimes \bar{P}^{(\kappa)} 
\end{eqnarray} 
where the second projector on the r.h.s. refers to the space spanned by the  
$\kappa$-mode SHFs,
\begin{eqnarray} \label{eq:SHF_projector}
    \bar{P}^{(\kappa)} =  \sum_{l,l^{\prime}} | {\Psi}^{(\kappa)}_{l^{\prime}} \rangle 
\,(\bm{\rho}^{(\kappa)})^{-1}_{l^{\prime}l}\, \langle      {\Psi}^{(\kappa)}_{l} |
\end{eqnarray}
Importantly, $\bar{P}^{(\kappa)}$ represents }
a projector onto a non-orthog\-onal basis, which includes the inverse of the 
overlap matrix. In the above, we used the definition
Eq.\ (\ref{eq:shf_overlap}), i.e., the SHF overlap coincides with the single-particle density matrix $\bm{\rho}^{(\kappa)}$.

To summarize, Eqs.\ (\ref{eq:projector_standard}), (\ref{eq:p0_projector}),
and (\ref{eq:pkappa_projector}) fully define the tangent space projection for
the MCTDH ansatz. {\color{black} We emphasize that, by construction, the
projectors $\mathcal{P}_{\kappa}({\Psi})$ are orthogonal to the projector
$\mathcal{P}_0({\Psi})$ and to each other.}

Anticipating the discussion of Sec.~\ref{sec:projector-splitting-eof}, the
novel projector-splitting algorithm will be shown to rely on a partitioning of
the ${\cal P}_\kappa (\Psi)$ projector of Eq.\ (\ref{eq:pkappa_projector})
into two components, ${\cal P}_\kappa (\Psi) = {\cal P}_\kappa^+ (\Psi) -
{\cal P}_\kappa^- (\Psi) = 1 \otimes \bar{P}^{(\kappa)} - P^{(\kappa)}\otimes
\bar{P}^{(\kappa)}$, see Eq.\ (\ref{split-projection}) below, permitting a new
partitioning of the equations of motion.

\subsection{MCTDH equations of motion}
\label{subsec:MCTDHeom}

The terms constituting $\mathcal{P}(\Psi)$ in
Eq.\ (\ref{eq:projector_standard}) give rise to the equations of the MCTDH
standard formulation, when applied to the TDSE as partial projections
according to Eq.\ (\ref{projected_TDSE}). Notably, we will consider the
conditions
\begin{eqnarray} 
\mathcal{P}_0(\Psi) \biggl[ \dot{\Psi}  - \frac{1}{{\imath \hbar}} H \Psi
\biggr] = 0 & ; &
\mathcal{P}_\kappa(\Psi) \biggl[ \dot{\Psi}  - \frac{1}{{\imath \hbar}} H \Psi
\biggr] =
0 \nonumber \\
\end{eqnarray} 
where
\begin{multline}
   \dot{\Psi}  - \frac{1}{\imath \hbar} H \Psi = \\
   \sum_{{J}} \; \dot{A}_{{J}} \; \Phi_{{J}} + \sum_{\kappa, l}  \;
   \dot{\varphi}_{l}^{(\kappa)} {\Psi}_{l}^{(\kappa)}  -
    \frac{1}{\imath \hbar} \sum_{{J}} \; {A}_{{J}} \; H \Phi_{{J}} 
\end{multline}
assuming the standard
MCTDH gauge that keeps the SPFs orthonormal during the propagation.

The $\mathcal{P}_0$ projection then yields
the standard differential equation for the $A$ vector,
\begin{equation} \label{eq:regMCTDH_coeff_diffequation}
   \dot{A}_{{I}} = \frac{1}{\imath \hbar} \sum_{{J}} \langle  \Phi_{{I}} | H | \Phi_{{J}} \rangle \; {A}_{{J}}
\end{equation}
whereas the projection along
$\mathcal{P}_{\kappa}(\Psi)$ returns the differential
equation for the SPFs of mode $\kappa$,
\begin{equation} \label{eq:regMCTDH_spf_diffequation}
\dot{\bm{\varphi}}^{(\kappa)} = 
\frac{1}{\imath \hbar}\left(1- P^{(\kappa)}  \right)
 (\bm{\rho}^{(\kappa)})^{-1} \; \langle \bm{H} \rangle^{(\kappa)} \; \bm{\varphi}^{(\kappa)}
\end{equation}
where $\bm{\varphi}^{(\kappa)}$ is the vector composed by the SPFs for mode $\kappa$ and 
$\langle \bm{H} \rangle^{(\kappa)}$ is the the usual mean-field potential matrix, given by 
\begin{equation}
   \langle {H} \rangle^{(\kappa)}_{jk}   = \langle {\Psi}_{j}^{(\kappa)} | H | {\Psi}_{k}^{(\kappa)} \rangle
\end{equation}
Next, we turn to the reformulation of the MCTDH equations
according to Ref.\ [\citenum{Lubich2015TimeDynamics}].

\section{The projector--splitting equations of motion}
\label{sec:projector-splitting-eof}

From the definition of the tangent-space projector
Eq.\ (\ref{eq:projector_standard}) and the resulting MCTDH equations, it is
clear that the SHF projector Eq.\ (\ref{eq:SHF_projector}) is at the origin of
the inverse of the density matrix appearing in
Eq.\ (\ref{eq:regMCTDH_spf_diffequation}). Hence, one can envisage an
{orthogonalizing transformation} in the SHF
space \cite{Lubich2015TimeDynamics}, such that the projector
$\bar{P}^{(\kappa)}$ of Eq.\ (\ref{eq:SHF_projector}) takes the alternative
form,
\begin{eqnarray}
    \bar{P}^{(\kappa)} =  \sum_{l} | \tilde{\Psi}^{(\kappa)}_{l} \rangle\langle \tilde{\Psi}^{(\kappa)}_{l} |
\end{eqnarray}
This concept is a key ingredient of the projector-splitting algorithm. As a
trade-off for the resulting simplification of the equations of motion of the
SPFs, the time-dependent transformation between non-orthogonal and
orthogonalized SHFs has to be taken into account. As will be shown below, this
can be conveniently achieved in terms of the splitting of the ${\cal P}_\kappa
(\Psi)$ projectors appearing in Eq.\ (\ref{eq:pkappa_projector}).

\subsection{SHF orthogonalization} 

First, we focus on the SHF orthogonalization and give a detailed description
of the new quantities introduced by this transformation. By construction, the
SHFs within the $\kappa$th subspace, ${\Psi}_{l}^{(\kappa)}$, are
non-orthogonal, and their overlap is given in terms of the reduced density
matrix $\bm{\rho}^{(\kappa)}$, see Eq.\ (\ref{eq:shf_overlap}). For the
purpose of the present discussion, we re-write the latter as follows,
\begin{equation} \label{eq:Overlap_vector_cuts}
    {\rho}^{(\kappa)}_{ll^{\prime}}
    = \sum_{{J}^{\kappa}} A_{{J}^{\kappa}_{l}}^\star
     A_{{J}^{\kappa}_{l^{\prime}}}
\end{equation}
where the vectors $A_{{J}^{\kappa}_{l^{\prime}}}$ were defined in Eq.\ (7).
In the following, we will interpret
$A_{{J}^{\kappa}_{l^{\prime}}}$ as a matrix composed of
$n_\kappa$ column vectors of length $\bar{n}_\kappa = n_1 \times \dots n_{\kappa-1}
\times  n_{\kappa+1} \dots \times n_f $ obtained by fixing the $\kappa$-th
index of the tensor. (In the tensor formulation, these vectors constitute a
matrix ${\boldsymbol{A}}^{(\kappa)}$ which is defined
as the $\kappa$-mode {\em matricisation} of the tensor $A$, see Appendix \ref{sec:matricisation-tensorisation}.)

Assuming that the $n_\kappa$ column vectors of $A_{{J}^{\kappa}_{l^{\prime}}}$ are linearly independent, we define a linear transformation
that brings them in orthonormal form, which we conveniently write as 
\begin{equation} \label{eq:QRdecomposition}
   A_{{J}^{\kappa}_{l}} = \sum_{l^{\prime}} {S}^{(\kappa)}_{ll^{\prime}} 
  Q^{(\kappa)}_{{J}^{\kappa}_{l^{\prime}}} 
\end{equation}
where $\bm{S}^{(\kappa)}$ is a lower triangular matrix and the tensor
$Q^{(\kappa)}$ is composed of orthogonal vectors in the sense that was
discussed above for the $A$ tensor. If we interpret both $A$ and
$Q^{(\kappa)}$ as matrices, with $f-1$ indices contained in a single
multi-index ${J}^{\kappa}$, we see that
Eq.\ (\ref{eq:QRdecomposition}) corresponds to a {QR decomposition} \cite{Golub}, {\it
i.e.}, the decomposition of the matrix ${\boldsymbol{A}}^{(\kappa)}$ as a product of an orthogonal matrix times
a triangular matrix,
{\color{black} 
\begin{equation} \label{eq:QRdecomposition_matrix}
 {\boldsymbol{A}}^{(\kappa)} = {\boldsymbol{Q}}^{(\kappa)}
 {\boldsymbol{S}}^{(\kappa)T}
\end{equation}
where ${\boldsymbol{S}}^{(\kappa)}$ is of dimension
$n_\kappa \times n_\kappa$ and ${\boldsymbol{Q}}^{(\kappa)}$ is of
dimension $\bar{n}_\kappa \times n_\kappa$.
}

From a numerical viewpoint, the QR decomposition is
generally stable and robust \cite{NoteQR,Golub}
and standard implementations with column pivoting \cite{Golub}} are adapted to situations where near-linear dependencies of the vectors extracted
 from $A$ occur (which is equivalent to the ill-conditioning of the density matrix, in light
of Eq. (\ref{eq:Overlap_vector_cuts})), {\color{black} see also 
Refs.\ [\citenum{Kloss2017ImplementationApproach,Golub}].}

Substituting the QR decomposition of Eqs.\ (\ref{eq:QRdecomposition})-(\ref{eq:QRdecomposition_matrix}) into the MCTDH {\it ansatz},
we can see that the $\bm{S}^{(\kappa)}$ matrix effectively gives rise to a
non-unitary transformation of the SPFs of the $\kappa$-th mode,
\begin{equation} \label{eq:MCTDH_QR_decomposition}
   \Psi =  \sum_{{J}} Q^{(\kappa)}_{{J}}
   \varphi^{(1)}_{j_1} \varphi^{(2)}_{j_2} {\dots} 
    \left[ \sum_{l} \varphi^{(\kappa)}_{l} {S}^{(\kappa)}_{l j_\kappa} \right] {\dots}
\end{equation}
In other words, $Q^{(\kappa)}_{{J}}$ is the tensor of the coefficients  
of an equivalent MCTDH expansion in which the $\kappa$-mode SPFs $\varphi^{(\kappa)}_{l}$ are substituted with {\em non-orthogonal} functions $\tilde{\varphi}^{(\kappa)}_{l} $
defined by the triangular matrix transformation $\bm{S}^{(\kappa)}$,
\begin{equation} \label{eq:SPF_nonorthogonaldefinition}
  \tilde{\bm{\varphi}}^{(\kappa)} = \bm{S}^{(\kappa)T} {\bm{\varphi}}^{(\kappa)}
\end{equation}
This transformation can also be understood as a QR decomposition if the
representation of the SPFs in the primitive representation is considered
as in Eq.\ (\ref{eq:tucker_format}). 

Within the transformed representation, the new SHFs are defined in accordance with the standard definition, namely
\begin{equation} \label{eq:SHF_orthogonaldefinition}
   \tilde{\Psi}^{(\kappa)}_{l} =
   \sum_{{J}^{\kappa}} Q^{(\kappa)}_{{J}^{\kappa}_{l}}  \;
 \prod_{n \neq \kappa}^{f}  \varphi_{j_n}^{(n)}  
\end{equation}
Importantly, the SHFs are now {\em orthonormal} by construction, because of the orthonormality of the ``vector cuts'' 
along $\kappa$ of ${\bm Q}^{(\kappa)}$, 
\begin{equation}
  \label{SHF_orthogonality}
  \langle \tilde{\Psi}^{(\kappa)}_{l} | \tilde{\Psi}^{(\kappa)}_{l^{\prime}} \rangle =
   \sum_{{J}^{\kappa}} Q_{{J}^{\kappa}_{l}}^{\star} \;
 Q_{{J}^{\kappa}_{l^{\prime}}} = \delta_{ll^{\prime}}
\end{equation}
In conclusion, an alternative SPF-SHF decomposition of the MCTDH wavefunction
has been constructed,
\begin{eqnarray} 
\Psi = \sum_{l} \tilde{\varphi}^{(\kappa)}_{l} \tilde{\Psi}^{(\kappa)}_{l} 
\label{psi_transformed}
\end{eqnarray}
which is analogous to the original representation (see Eq. \ref{eq:spf_shf_decomposition}) with the difference that
the SHFs are now orthogonal and the SPFs are not. 

Finally, substituting the QR decomposition of Eqs.\ (\ref{eq:QRdecomposition})-(\ref{eq:QRdecomposition_matrix}) into the
expression of the reduced density matrix of Eq.\ (\ref{eq:Overlap_vector_cuts}) we obtain
\begin{equation}
  \label{rho_S}
 \bm{\rho}^{(\kappa)} = \bm{S}^{(\kappa)} \bm{S}^{(\kappa)\dagger}
\end{equation}
{\it i.e.}, the reduced density matrix factorizes in the form of a Cholesky
decomposition \cite{Golub}.
From this, we can better understand that the uniqueness of the matrix
$\bm{S}^{(\kappa)}$ is closely
connected to the invertibility of $\bm{\rho}^{(\kappa)}$, as the Cholesky decomposition is unique for
strictly positive-definite matrices.

\subsection{Splitting of subspace projections} 

Using the new SHFs $\tilde{\Psi}^{(\kappa)}_{l}$ of
Eq.\ (\ref{eq:SHF_orthogonaldefinition}), we can now express the operator $\mathcal{P}_{\kappa}(\Psi)$ as
\begin{equation} \label{eq:new_projector}
\mathcal{P}_{\kappa}(\Psi) = \left( {\color{black} {1}} - \sum_i | \varphi^{(\kappa)}_{i} \rangle \langle
\varphi^{(\kappa)}_{i} |   \right) \otimes  \sum_{l} | \tilde{\Psi}^{(\kappa)}_{l} \rangle \langle
\tilde{\Psi}^{(\kappa)}_{l} |
\end{equation}
where the orthogonality of the SHFs is made evident by the disappearance of the overlap
matrix from the projector. Eq.\ (\ref{eq:new_projector}) is a hybrid
representation where we keep the SPFs in their orthogonal form.
As will become clear in the following, this is motivated by the
fact that we will construct a suitable subprojection that singles out the
time derivative of $\bm{S}^{(\kappa)}$. 

We further divide each of the projectors $\mathcal{P}_{\kappa}(\Psi)$
of Eq.\ (\ref{eq:projector_standard}) into two new projectors,
splitting $(1 - \sum_i | \varphi^{(\kappa)}_{i} \rangle \langle \varphi^{(\kappa)}_{i} |)$ into two components:
\begin{subequations}
\begin{eqnarray}
     \mathcal{P}^{+}_{\kappa}(\Psi) =& {1}  &\otimes  \sum_{l} | \tilde{\Psi}^{(\kappa)}_{l} \rangle
\langle \tilde{\Psi}^{(\kappa)}_{l} |  \label{Pplus} \\
   \mathcal{P}^{-}_{\kappa}(\Psi) =& \sum_{i} | \varphi^{(\kappa)}_{i} \rangle \langle
\varphi^{(\kappa)}_{i} |   &\otimes 
       \sum_{l} | \tilde{\Psi}^{(\kappa)}_{l} \rangle \langle \tilde{\Psi}^{(\kappa)}_{l} | \label{Pminus}
\end{eqnarray}
\end{subequations}
such that the overall projector now reads
\begin{equation}
  \label{split-projection}
   \mathcal{P}(\Psi) = \mathcal{P}_0(\Psi) + \sum_{\kappa = 1}^{f} \biggl( \mathcal{P}^{+}_{\kappa}(\Psi) -
\mathcal{P}^{-}_{\kappa}(\Psi) \biggr) 
\end{equation}

These newly defined projection operators give rise to a different formulation
of the differential equations in the $\kappa$-subspaces, which is equivalent
to the original MCTDH formulation but makes direct use of the new SPF-SHF
decomposition. Meanwhile, Eq. (\ref{eq:regMCTDH_coeff_diffequation}) remains
unchanged, since the projector $\mathcal{P}_0(\Psi)$ is left unchanged by the
projector splitting.

From the action of $\mathcal{P}^{+}_{\kappa}(\Psi)$ on the time-dependent Schr\"o\-dinger equation, we now
obtain the following expression for the propagation of the non-orthogonal SPFs,
\begin{equation} \label{eq:newMCTDH_spf_diffequation}
   \dot{\tilde{\bm{\varphi}}}^{(\kappa)} = \frac{1}{\imath \hbar}
   \langle \tilde{\bm{H}} \rangle^{(\kappa)} \tilde{\bm{\varphi}}^{(\kappa)}
\end{equation}
{as detailed in Appendix \ref{sec:projector_splitting-eom}}. In
Eq.\ (\ref{eq:newMCTDH_spf_diffequation}), the new mean-field potential $\langle \tilde{\bm{H}} \rangle^{(\kappa)}$ is defined by integrating over 
the orthogonalized SHFs,
\begin{equation} \label{eq:new_mean_field}
   \langle \tilde{{H}} \rangle^{(\kappa)}_{ll^{\prime}} =
   \langle \tilde{\Psi}^{(\kappa)}_{l} | H | \tilde{\Psi}^{(\kappa)}_{l^{\prime}} \rangle
\end{equation}
The advantage of Eq.\ (\ref{eq:newMCTDH_spf_diffequation}) over Eq.\ (\ref{eq:regMCTDH_spf_diffequation})
is evident: the reduced density matrix has been incorporated in the expression for the SPFs and the
evaluation of the expression no longer requires the inversion of a potentially singular matrix.

The price to pay for this transformation is that we are now dealing with non--orthogonal SPFs,
and an additional differential equation appears which is generated by 
the projector
$\mathcal{P}^{-}_{\kappa}(\Psi)$. Notably, we obtain 
\begin{equation} \label{eq:newMCTDH_S_diffequation}
   \dot{{S}}^{(\kappa)}_{ij} =    \frac{1}{\imath \hbar} \langle {\varphi}^{(\kappa)}_{i} |
   \sum_{k} \left(    \langle \tilde{\bm{H}} \rangle^{(\kappa)}
       {\bm{S}}^{(\kappa)T}   \right)_{jk} | {\varphi}^{(\kappa)}_{k} \rangle
\end{equation}
{where we again refer to Appendix \ref{sec:projector_splitting-eom}
for details of the derivation.} The above expression involves matrix elements of the mean-field operators of Eq.\ (\ref{eq:new_mean_field}) multiplied
by the transformation matrix. 

The combination of Eq.\ (\ref{eq:regMCTDH_coeff_diffequation}),
Eq.\ (\ref{eq:newMCTDH_spf_diffequation}), and
Eq.\ (\ref{eq:newMCTDH_S_diffequation}), which define the new equations of
motion, necessitate toggling between the two SPF-SHF representations of
Eq.\ (\ref{eq:spf_shf_decomposition}) and Eq.\ (\ref{psi_transformed}). As
will be further discussed below, this is achieved by the QR decomposition
steps of Eq.\ (\ref{eq:QRdecomposition_matrix}) and
Eq.\ (\ref{eq:SPF_nonorthogonaldefinition}).

The above equations have been obtained with the standard gauge condition
for the original SPFs,
\begin{equation}
      \langle \dot{\varphi}^{(\kappa)}_{i} | {\varphi}^{(\kappa)}_{j} \rangle = 0 \;\; \forall\, i,j
\end{equation}
along with an additional gauge condition for the orthonormalized SHFs,
\begin{equation}
      \langle \dot{\tilde{\Psi}}^{(\kappa)}_{k} |
      {\tilde{\Psi}}^{(\kappa)}_{l} \rangle = 0
      \;\; \forall\, k,l
      \label{gauge_SHF}
\end{equation}
This additional gauge is equivalent to the condition \break
$\dot{\boldsymbol{Q}}^{(\kappa)\dagger} {\boldsymbol{Q}}^{(\kappa)}~=~0$, as
follows from Eq.\ (\ref{SHF_orthogonality}), and guarantees that the new SHFs
remain orthonormal during the propagation. (Alternatively, the presence of an
additional gauge condition can be taken to arise because of the QR
decomposition of the $A$ coefficients according to
Eqs.\ (\ref{eq:QRdecomposition}) and (\ref{eq:QRdecomposition_matrix}), which
necessitates an additional gauge \cite{MeyerWang}.) 

\subsection{Integration scheme}
\label{splitting-integrator}

The implementation of the above equations
Eq.\ (\ref{eq:regMCTDH_coeff_diffequation}),
Eq.\ (\ref{eq:newMCTDH_spf_diffequation}), and
Eq.\ (\ref{eq:newMCTDH_S_diffequation}), as described in
Refs.\ [\citenum{Lubich2015TimeDynamics,Kloss2017ImplementationApproach}], is
detailed in Appendix~\ref{sec:projector-splitting-integration-scheme}.
{\color{black} Here, we give some introductory remarks.}

From a general perspective, the projector--splitting integrator follows the idea of a second-order
scheme which is known as {Strang splitting} \cite{Lubich2008FromAnalysis} in the mathematical literature. Well-known examples
of this type of integrators in the physical sciences are the popular {velocity-Verlet method}
in classical molecular dynamics \cite{Tuckerman2010StatisticalSimulations} and the {second-order split-operator method} in
quantum dynamics \cite{Feit1982SolutionMethod}. These algorithms, each in its own appropriate formalism, share
the use of the {symmetric Trotter expansion} of the exponential, {\it i.e.}, the
approximation
\begin{equation}
\label{strang_splitting}  
\text{e}^{(A+B) \, \delta t} \sim \text{e}^{\frac{1}{2} A \, \delta t } \text{e}^{B \, \delta t}
                              \text{e}^{\frac{1}{2} A \, \delta t }
\end{equation}
These type of integrators have attractive general properties, including
unitarity and the preservation of the underlying symplectic structure of the space in which the solution evolves. In our case, the
approximated exponential is the formal solution of Eq.\ (\ref{eq:orthogonal_proj_derivative}) over a short time interval which reads as the
propagator in the MCTDH tensor--product space
{\color{black} \begin{equation}
  \Psi(\delta t) = {\cal U}( \delta t ) \Psi(0) = \exp \left( -\frac{\imath}{\hbar} {\cal
 P}(\Psi) H \, \delta t \right)  \Psi(0)
\end{equation}
and is approximated according to the projector splitting scheme of
Eq.\ (\ref{eq:projector_standard}) combined with
Eq.\ (\ref{strang_splitting}),
\begin{equation}
  {\cal U}( \delta t )= \biggl( \prod_{\kappa=1}^f {\cal U}_\kappa ( \delta t/2 )  \biggr)
  {\cal U}_0 ( \delta t ) \biggl( \prod_{\kappa=1}^f {\cal U}_\kappa ( \delta t/2 )  \biggr)
\end{equation}
with   
\begin{equation}
  {\cal U}_0 ( \delta t )= \text{e}^{-{\imath} {\cal P}_0(\Psi) H \, \delta
  t/\hbar} \quad , \quad  {\cal U}_\kappa ( \delta t ) = \text{e}^{-{\imath}
  {\cal P}_\kappa(\Psi) H \, \delta t/\hbar}
\end{equation}

In light of the above,} each integration interval $\delta t$ is constructed by
a sequence of three steps: (i) propagation of the SPFs during a half-step
$\frac{1}{2}\delta t$, (ii) propagation of the $A$ vector during a full step
$\delta t$ and (iii) propagation of the SPFs during a second half-step
$\frac{1}{2}\delta t$. While the second step is constructed as in the standard
MCTDH scheme, the first and {\color{black} third} steps are based on
Eqs.\ \ (\ref{eq:newMCTDH_spf_diffequation})
and\ (\ref{eq:newMCTDH_S_diffequation}) instead of Eq.
(\ref{eq:regMCTDH_spf_diffequation}) {\color{black} in the new algorithm}.
{\color{black} Similarly to the constant mean field (CMF) integration
scheme \cite{Beck1997} of MCTDH, the mean fields are kept constant during the
SPF integration intervals.}

{Two key issues that need to be considered in the implementation of
the algorithm are as follows:}

First, since the propagation of the $A$ coefficients
  relies on the regular SPFs ${\varphi}^{(\kappa)}_{l}$, these need to be
  reconstructed after each SPF integration half-step
  from the propagated non-orthogonal SPFs $\tilde{\varphi}^{(\kappa)}_{l}$ and
  the transformation matrix ${\bm{S}}^{(\kappa)}$.
  If this was done by an inversion of the ${\bm{S}}^{(\kappa)}$ matrix,
${\bm{\varphi}}^{(\kappa)} = (\bm{S}^{(\kappa)T})^{-1}
\tilde{\bm{\varphi}}^{(\kappa)}$ according to
Eq.\ (\ref{eq:SPF_nonorthogonaldefinition}), issues about
ill--conditioning would arise, in exactly the same way as for the reduced
density matrix. The present algorithm circumvents this problem
by a QR decomposition of the propagated SPFs $\tilde{\varphi}^{(\kappa)}_{l}$
according to Eq.\ (\ref{eq:SPF_nonorthogonaldefinition}),
$\tilde{\bm{\varphi}}^{(\kappa)} = \bm{S}^{(\kappa)T}
{\bm{\varphi}}^{(\kappa)}$. By definition, the resulting
regular SPFs ${\varphi}^{(\kappa)}_{l}$ are orthogonal.

\begin{figure*}
  \centering \includegraphics[width=0.9\textwidth]{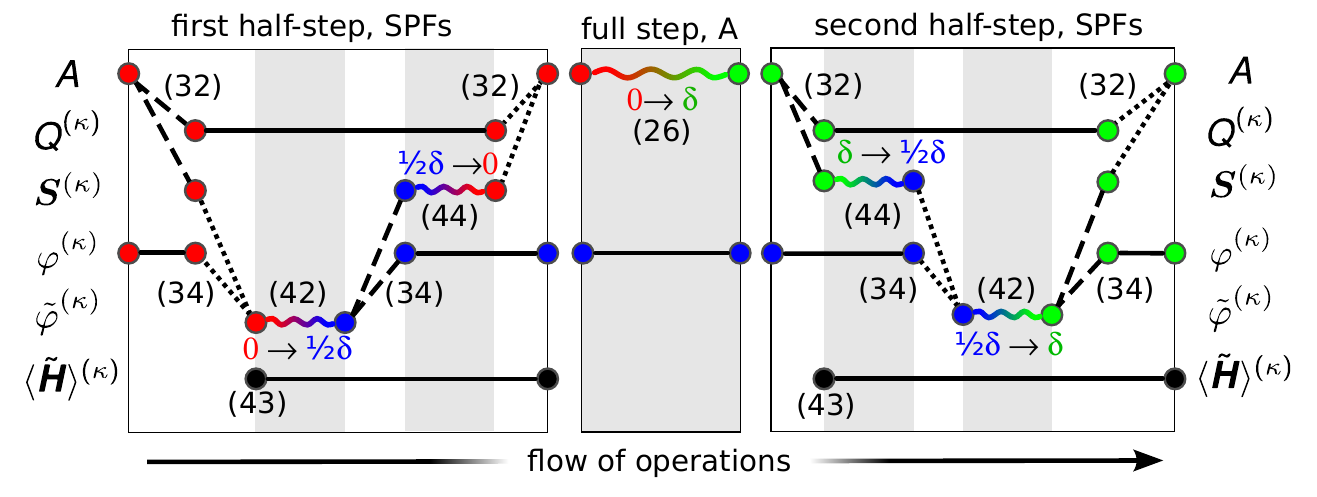}
   \caption{{\color{black} Schematic representation  of the sequence of operations in the
   integration algorithm.
   Each line shows how the
   relevant quantities defined in the text are updated in the three steps of the
   algorithm. QR decompositions are indicated by dashed lines, ${\bm{S}}^{(\kappa)}$ multiplication by
   dotted lines and time integration by wiggly lines. Numbers in parentheses refer to the
   relevant equations in the text.
   Colors are used to indicate
   different times (red for $t_0 = 0$, blue for $t_0 + \frac{1}{2}\delta = \frac{1}{2}\delta$ and
      green for $t_0 + \delta = \delta$) where the relevant quantities are evaluated. Note that the
   mean fields are kept constant (at their values at times $t = t_0 = 0$ and
   $t = t_0 + \delta = \delta$, respectively)
   during the first and second half-steps of the SPF propagation.}}
   \label{fig:integrationscheme}
\end{figure*}

Second, the propagation of the
   $A$ coefficients and the SPFs has to be consistent,
  in the sense that the time evolution described by the $A$ coefficients is
  not ``repeated'' by the SPFs and {\em vice versa}. This property, which is
  visible in the standard MCTDH equations of motion,
  Eq.\ (\ref{eq:regMCTDH_spf_diffequation}), in terms of the $(1 -
  P^{(\kappa)})$ projector, is now encoded in the ${\bm{S}}^{(\kappa)}$
  evolution. However, a complication arises from the fact that
  ${\bm{S}}^{(\kappa)}$ is not only updated according to
  Eq.\ (\ref{eq:newMCTDH_S_diffequation}), but also by QR decomposition of the
  $A$ coefficients, Eq.\ (\ref{eq:QRdecomposition}), and of the transformed
  SPFs, Eq.\ (\ref{eq:SPF_nonorthogonaldefinition}). In particular, if the
  updated SPFs are generated by QR decomposition following propagation of the
  $\tilde{\bm{\varphi}}^{(\kappa)}$, using $\tilde{\bm{\varphi}}^{(\kappa)} =
  \bm{S}^{(\kappa)T} {\bm{\varphi}}^{(\kappa)}$ as explained above, the
  updated $\bm{S}^{(\kappa)T}$ is not identical to $\bm{S}^{(\kappa)T}$ as
  obtained by time propagation according to
  Eq.\ (\ref{eq:newMCTDH_S_diffequation}). This needs to be corrected for by
  additional (back-) propagation steps \cite{Lubich2015TimeDynamics} of
  $\bm{S}^{(\kappa)T}$ serving as a ``gauge correction'', as detailed
  {\color{black} in Appendix
    \ref{sec:projector-splitting-integration-scheme}.}

These considerations lead to an algorithm \cite{Lubich2015TimeDynamics,Kloss2017ImplementationApproach}
involving a simultaneous backward and forward-in-time propagation accompanied
by two QR decompositions per step, as depicted in
Fig.\ (\ref{fig:integrationscheme}) and further detailed in Appendix
\ref{sec:projector-splitting-integration-scheme}. 
Here, the transformation ${\bm{S}}^{(\kappa)}$ is handled as an
auxiliary quantity that is continuously updated during the algorithm.
As underscored in Ref.\ [\citenum{Lubich2015TimeDynamics}], the algorithm
does not use any pre-determined gauge as in the standard MCTDH formulation,
but adapts the gauge {\em via} QR decompositions (or, alternatively, singular
value decompositions \cite{Lubich2015TimeDynamics}).

\section{Summary and conclusions}
\label{sec:summary}

The aim of this article is to make some recent results obtained in the mathematics community more
accessible to a chemical physics audience, specifically in the context of the 
new projector-splitting integrator for MCTDH developed by Lubich \cite{Lubich2015TimeDynamics}
and recently implemented by Kloss et al. \cite{Kloss2017ImplementationApproach}.
To this end, several aspects have been highlighted in the present work: First, the
derivation of the standard MCTDH equations from the tangent space projection
Eq.\ (\ref{eq:projector_standard}), as previously derived in a mathematical
context in 
Ref.\ [\citenum{Koch2010DynamicalApproximation}].
Second, the complementary derivation
of the modified MCTDH equations presented in Ref.\ [\citenum{Lubich2015TimeDynamics}],
from the perspective of a suitable splitting of the tangent space projection
for MCTDH, according to Eq.\ (\ref{split-projection}).  
Third, the concept of orthogonalizing the SHFs to formally eliminate the
inverse of the single-particle density matrix from the equations of motion.
Finally, the structure of the algorithm designed by Lubich \cite{Lubich2015TimeDynamics,Kloss2017ImplementationApproach} to make the
new propagation scheme efficient.

The tangent space projection of the
MCTDH wave\-function \cite{Koch2010DynamicalApproximation} (Sec.\ \ref{sec:tangent-space-projection}),
is not a common tool
so far in the derivation of variational equations from the viewpoint of the chemical
physics community. We believe that this perspective can be most useful in
understanding the structure of the variational equations and designing new
approximation schemes. {\color{black} Recent work in the area of matrix
product states \cite{Haegeman16,Haegeman13,Haegeman11}
underscores the usefulness of this approach. }

The projector splitting algorithm as described in
Sec.\ \ref{splitting-integrator} and Appendix \ref{sec:projector-splitting-integration-scheme} generally leads to a robust
propagation \cite{Kloss2017ImplementationApproach} and circumvents the
regularization procedure of standard MCTDH. However, the QR decomposition
steps are also affected by singularities of the density matrix, since the QR
decomposition is non-unique for rank-deficient matrices ${\bm{S}}^{(\kappa)}$
and, hence, ${\boldsymbol{\rho}}^{(\kappa)}$, see Eq.\ (\ref{rho_S}).
Therefore, the propagation does have some dependence on how the QR algorithm
treats the rank-deficient case \cite{Golub}, see the discussion of
Ref.\ [\citenum{Kloss2017ImplementationApproach}]. Whether or not the
algorithm handles the initially unoccupied SPFs in an advantageous way,
especially as compared with the construction of optimal unoccupied
SPFs \cite{Manthe2015TheRevisited}, is currently a matter of
debate \cite{MeyerWang} and needs to be further investigated in numerical
studies.

\section{Acknowledgments}
It is a great pleasure to dedicate this paper to Wolfgang Domcke on
the occasion of his 70th birthday.
We thank the German-Israeli Foundation for Scientific Research and Development for support of this project under grant number GIF I-1337-302.5/2016. 
M.B. gratefully acknowledges fellowship support by the Alexander von Humboldt
Foundation.
{We thank H.-D. Meyer for discussions and suggestions in
the context of the integration scheme.}

\appendix

\section*{Appendices}

\setcounter{equation}{0}
\renewcommand{\theequation}{\Alph{section}.\arabic{equation}}
\renewcommand*{\thesection}{\Alph{section}}

\section{Basic notions of tensor algebra}
\label{sec:matricisation-tensorisation}

In tensor language, the MCTDH {\it ansatz} is known as Tucker
format \cite{Koch2010DynamicalApproximation,Lubich2013DynamicalTensors,Bachmayr2016TensorEquations},
in which $\Psi$ is decomposed in terms of a core--tensor (the $A$--vector of
conventional MCTDH) and a rectangular matrix per physical dimension (the
$U^{(\kappa)}_{i_\kappa j_\kappa}$ matrix of Eq.\ (\ref{eq:tucker_format}))
representing the change of basis from the primitive grid to the SPFs. Other
alternative decompositions have been explored both in the chemical physics and
mathematical literature, notably relating to the ML-MCTDH scheme that
corresponds to a hierarchical Tucker decomposition. We refer to
Ref.\ [\citenum{Bachmayr2016TensorEquations}] for a topical survey of the
field.

To write the standard tensor operations in concise form,
the  formalism which is used in Refs.\ [\citenum{Lubich2015TimeDynamics,Kloss2017ImplementationApproach}]
makes extensive use of the concepts of {\em matricisation} and {\em tensorisation}, which are a key point
in translating the equations to common MCTDH notation.

We illustrate these concepts for operators in a sum-of-products (SOP) form,
where the application of the operator can be split into the sequential application of smaller
matrices, as has been recognized early on in the context of MCTDH \cite{Manthe1992Wave-PacketNOCl,Beck2000TheWavepackets}.

In the tensor formalism, this type of operation is explicitly written by means
of the matricisation of the tensor. When an operator with matrix
representation $\bm{O}^{(\kappa)}$
acting specifically on mode $\kappa$ is applied
to a tensor $A$, it is convenient to recast the tensor in matrix form
$\bm{A}^{(\kappa)}$ such that its rows are labeled by the index of the
primitive basis of mode $\kappa$. In formulas, the {\it $\kappa$-mode
matricisation} $\bm{A}^{(\kappa)}$ is defined as
\begin{equation}
    {A}^{(\kappa)}_{j_\kappa{J}^{\kappa}} = A_{{J}}
\end{equation}
with the definition of multi--indices as described in the main text.
With $\bm{A}^{(\kappa)}$,
the action of the operator $\hat{O}^{(\kappa)}$
can be computed using conventional matrix multiplication,
\begin{equation}
    \bm{A}^{(\kappa)\prime} = \bm{O}^{(\kappa)} \bm{A}^{(\kappa)}
\end{equation}
The tensor corresponding to the resulting matrix can be reconstructed by {\it
tensorization}, {\it i.e.} by reordering the components and labeling them by
the usual multi-index ${J}$.

The sequential operations consisting in (i) $\kappa$-mode matricisation (ii) matrix multiplication with a
$\kappa$-mode matrix and (iii) tensorization are written concisely as
\begin{equation} \label{eq:tensor_matrix_product}
   A^{\prime} = A \times_\kappa \bm{O}^{(\kappa)}
\end{equation}
or, with a full specification of indices,
\begin{equation} \label{eq:tensor_matrix_product_expl}
    {A}^{(\kappa)\prime}_{{J}^{\kappa}_{l}} = \sum_{j_\kappa}
    \bm{O}^{(\kappa)}_{l,j_\kappa} A_{{J}}
\end{equation}

By repeated application of this definition, the action of an operator in product form, 
$ \prod_\kappa \bm{O}^{(\kappa)} $, is written as
\begin{equation} \label{eq:repeated_tensor_matrix}
    A^{\prime} = A \times_{\kappa=1}^{f}
    \bm{O}  = A \times_{1} \bm{O}^{(1)} \times_{2} \bm{O}^{(2)} \dots
                \times_{f} \bm{O}^{(f)}
\end{equation}
which again can be explicitly written with a full specification of the indices as
\begin{equation}
    A^{\prime}_{{L}}  = \sum_{j_1}
    \dots \sum_{j_f} \bm{O}^{(1)}_{l_1,j_1}  \ldots
    \bm{O}^{(f)}_{l_f,j_f} \; A_{{J}}
\end{equation}

The reader should note that the operations defined above can be used to define not only the
operator matrix elements, but also the tensor decomposition in Tucker form.
In fact, by using the definition of Eq. (\ref{eq:repeated_tensor_matrix}), the MCTDH {\em ansatz} in
tensor form, i.e., Eq.\ (\ref{eq:tucker_format}), can be immediately recognized as
\begin{equation}
    \Psi = A \times_{\kappa=1}^{f} \bm{U}^{(\kappa)}
\end{equation}
where $A$ is the core--tensor and $\bm{U}^{(\kappa)}$ the representations of
the SPFs in the primitive basis.

By a full specification of the indices and by removing the primitive basis projection, all
equations of Refs.\ [\citenum{Lubich2015TimeDynamics,Kloss2017ImplementationApproach}] can be cast in 
standard MCTDH form.
We consider as another instance the definition of the SHFs
\begin{equation}
  {\Psi}^{(\kappa)} =
  \bm{A}^{(\kappa)} \bigotimes_{n \neq \kappa}^{f} {\bm{U}}^{(\kappa)T}
\end{equation}
The matrix ${\Psi}^{(\kappa)}$
has dimensions $n_\kappa \times (\prod_{i \neq \kappa} N_i ) $
($N_i$ being the dimension of the $i$-th primitive grid) and its elements are given by
\begin{equation}
{\Psi}^{(\kappa)}_{ l \, {I}^{\kappa} }
= \sum_{{J}^{\kappa}} \bm{A}^{(\kappa)}_{l \, {J}^{n} } \;
 \prod_{n \neq \kappa}^{f} \bm{U}^{(n)}_{i_n\, j_n}
\end{equation}
where we have substituted the expression for the Kronecker matrix product.
Again, by substituting the expression for the SPF expansion on the primitive basis we obtain
\begin{equation}
 {\Psi}^{(\kappa)}_{l \, {I}^{\kappa} } =
  \sum_{{J}^{\kappa}} A_{{J}^{\kappa}_{l}}  \;
 \prod_{n \neq \kappa}  \langle \chi_{i_n}^{(n)} | \varphi_{j_n}^{(n)} \rangle
\end{equation}
which is recognized as the primitive basis representation of the $l$-th SHF for mode $\kappa$
\begin{equation}
  {\Psi}^{(\kappa)}_{l \, {I}^{\kappa} } = \langle \chi_{i_1}^{(1)} \dots \chi_{i_{\kappa-1}}^{(\kappa-1)}
\chi_{i_{\kappa+1}}^{(\kappa+1)} \dots  |
\psi_{l}^{(\kappa)} \rangle
\end{equation}

\section{Projector splitting equations of motion}
\label{sec:projector_splitting-eom}

Here, we provide details of the derivation of the equations of motion
Eq.\ (\ref{eq:newMCTDH_spf_diffequation}) and
Eq.\ (\ref{eq:newMCTDH_S_diffequation}) for the quantities
$\tilde{\bm{\varphi}}^{(\kappa)}$ and ${\bm{S}}^{(\kappa)}$.
By analogy with Sec.\ \ref{subsec:MCTDHeom}, we consider the following equations for the
projectors $\mathcal{P}^{\pm}_{\kappa}(\Psi)$:
\begin{eqnarray}
  \mathcal{P}^\pm_{\kappa}(\Psi) \biggl[ \dot{\Psi}  - \frac{1}{\imath \hbar}
  H \Psi \biggr] = 0
\end{eqnarray} 
where $\dot{\Psi}$ is conveniently expressed in the form
of Eq.\ (\ref{psi_transformed}) such that 
\begin{multline}
   \dot{\Psi}  - \frac{1}{\imath \hbar} H \Psi = \\
   \sum_{l} \tilde{\varphi}^{(\kappa)}_{l}
   \dot{\tilde{{\Psi}}}^{(\kappa)}_{l} +
   \sum_{l} \dot{\tilde{\varphi}}^{(\kappa)}_{l}
   \tilde{\Psi}^{(\kappa)}_{l}
   - \frac{1}{\imath \hbar} \sum_{l} \;
   H  \tilde{\varphi}^{(\kappa)}_{l}
   \tilde{\Psi}^{(\kappa)}_{l} 
\end{multline}
From the explicit form of the projector $\mathcal{P}^+_{\kappa}(\Psi)$ in
Eq.\ (\ref{Pplus}) in conjunction with the additional gauge condition
Eq.\ (\ref{gauge_SHF}) for the orthogonalized SHFs, we obtain
\begin{eqnarray}
  \label{Pplus_derivation}
  \mathcal{P}^+_{\kappa}(\Psi)&& \hspace*{-0.8cm} \sum_{l} \biggl(
   \tilde{\varphi}^{(\kappa)}_{l} \dot{\tilde{\Psi}}^{(\kappa)}_{l}
   + \dot{\tilde{\varphi}}^{(\kappa)}_{l} \tilde{\Psi}^{(\kappa)}_{l} 
  - \frac{1}{\imath \hbar}  H \tilde{\varphi}^{(\kappa)}_{l} \tilde{\Psi}^{(\kappa)}_{l} 
   \biggr) \nonumber \\
   \mbox{} & = & \sum_{j} 
     \tilde{\Psi}^{(\kappa)}_{j}  \biggl( 
     \dot{\tilde{\varphi}}^{(\kappa)}_{j} 
     - \frac{1}{\imath \hbar} \sum_{l}
     \langle \tilde{{H}} \rangle^{(\kappa)}_{jl}
     \tilde{\varphi}^{(\kappa)}_{l} \biggr)  = 0
     \nonumber \\
\end{eqnarray}
with the mean-field Hamiltonian matrix element
$\langle \tilde{{H}} \rangle^{(\kappa)}_{jl} = \langle
\tilde{\Psi}^{(\kappa)}_{j} \vert H \vert \tilde{\Psi}^{(\kappa)}_{l} \rangle$.
Eq.\ (\ref{Pplus_derivation}) immediately yields Eq.\ (\ref{eq:newMCTDH_spf_diffequation}).

Turning to the projector $\mathcal{P}^-_{\kappa}(\Psi)$
in Eq.\ (\ref{Pminus}), we obtain
\begin{eqnarray}
  \label{Pminus_derivation}
  \mathcal{P}^-_{\kappa}(\Psi)&& \hspace*{-0.8cm} \sum_{l} \biggl(
   \tilde{\varphi}^{(\kappa)}_{l} \dot{\tilde{\Psi}}^{(\kappa)}_{l}
   + \dot{\tilde{\varphi}}^{(\kappa)}_{l} \tilde{\Psi}^{(\kappa)}_{l} 
  - \frac{1}{\imath \hbar}  H \tilde{\varphi}^{(\kappa)}_{l} \tilde{\Psi}^{(\kappa)}_{l} 
   \biggr) \nonumber \\
  \mbox{} & = & \sum_i \sum_{j} {\varphi}^{(\kappa)}_{i}
  {\tilde{\Psi}}^{(\kappa)}_{j} \biggl(
  \dot{S}^{(\kappa)T}_{li} \nonumber \\
  \mbox{} & & \hspace*{0.4cm}- \frac{1}{\imath \hbar}
  \sum_{l,k} \langle {\varphi}^{(\kappa)}_{i} \vert
  \langle \tilde{{H}} \rangle^{(\kappa)}_{jl}
  {S}^{(\kappa)T}_{lk} \vert {\varphi}^{(\kappa)}_{k}\rangle \biggr) = 0
  \nonumber \\
\end{eqnarray}
which directly results in Eq.\ (\ref{eq:newMCTDH_S_diffequation}).
In the second line of Eq.\ (\ref{Pminus_derivation}), we again used
the additional gauge condition
Eq.\ (\ref{gauge_SHF}), along with the relation $\langle
{\varphi}^{(\kappa)}_{i} \vert \dot{\tilde{\varphi}}^{(\kappa)}_{l} \rangle =
\dot{S}_{li}^{(\kappa)T}$,
which follows from Eq.\ (\ref{eq:SPF_nonorthogonaldefinition})
together with the standard gauge condition $\langle {\varphi}^{(\kappa)}_{i}
\vert \dot{\varphi}^{(\kappa)}_{j} \rangle = 0$.

\section{The projector-splitting integration scheme}
\label{sec:projector-splitting-integration-scheme}

{As shown in Fig.\ (\ref{fig:integrationscheme}) and summarized in
Sec.\ (\ref{splitting-integrator})},
the algorithm described in Refs.\
[\citenum{Lubich2015TimeDynamics,Kloss2017ImplementationApproach}] comprises a
sequence of three steps:
\begin{enumerate}
 \item a half-step for each of the $f$ modes, with the propagation of the SPFs and of the
${\bm{S}}^{(\kappa)}$ matrix (Eq.\ (\ref{eq:newMCTDH_S_diffequation}) and Eq.\
(\ref{eq:newMCTDH_spf_diffequation})),
 \item a full step for the $A$ vector (Eq.\ \ref{eq:regMCTDH_coeff_diffequation}),
 \item another half-step for the SPFs and the ${\bm{S}}^{(\kappa)}$ matrix.
\end{enumerate}

\vspace*{0.2cm}

In detail, the first step is carried out as follows:
\begin{itemize}[wide, labelwidth=!, labelindent=0pt]
 \item QR decomposition of the $A$ vector (Eq.\ (\ref{eq:QRdecomposition_matrix}))
       and construction of the corresponding
       non--orthogonal SPFs $\tilde{\bm{\varphi}}^{(\kappa)}$ (Eq.\ (\ref{eq:SPF_nonorthogonaldefinition}))
       and mean--field potentials (Eq.\ (\ref{eq:new_mean_field})) at time $t
       = t_0$.
 \item forward integration of Eq.\ (\ref{eq:newMCTDH_spf_diffequation}) for a half time--step $\delta$,
       to obtain new values of the non--orthogonal SPFs at time $t = t_0 +
       \frac{1}{2} \delta$:
 \begin{eqnarray}
   \Delta \tilde{\bm{\varphi}}^{(\kappa)} =
   \int_{t_0}^{t_0+\delta/2} dt'\, {f}_{\varphi}
   (\tilde{\bm{\varphi}}^{(\kappa)} (t') \ \vert
   \ \langle \tilde{\bm{H}} \rangle^{(\kappa)} (t_0) )
   \nonumber
 \end{eqnarray}
 with ${f}_\varphi( \ \cdot \ \vert \ \cdot \ )$ referring to the r.h.s.\ of
 Eq.\ (\ref{eq:newMCTDH_spf_diffequation}),
 where the left argument specifies the quantity which is integrated, while the
 right argument specifies the quantities that are held constant at a certain time
 ({\em cf.}\ Ref.\ [\citenum{Beck1997}] for a similar notation in the context
 of the constant mean field (CMF)
 integrator of MCTDH).
 \item QR decomposition of the updated non--orthogonal SPFs (Eq.\
       (\ref{eq:SPF_nonorthogonaldefinition})) to obtain values of the
       orthonormal SPFs ${\bm{\varphi}}^{(\kappa)}$ and of
       the matrix ${\bm{S}}^{(\kappa)}$ at time
       $t = t_0 + \frac{1}{2} \delta$.
 \item backward integration of Eq.\ (\ref{eq:newMCTDH_S_diffequation}) for a half time--step
       to obtain a $t_0$ value of
       ${\bm{S}}^{(\kappa)}$:
       \begin{eqnarray}
            \hspace*{-0.7cm}\Delta {\bm{S}}^{(\kappa)} = \int_{t_0+\delta/2}^{t_0} dt'\, {f}_S
             ({\bm{S}}^{(\kappa)}(t') \ \vert \ \langle \tilde{\bm{H}}
             \rangle^{(\kappa)} (t_0),
             {\bm{\varphi}}^{(\kappa)}(t_0+\frac{\delta}{2}) ) \nonumber
       \end{eqnarray}
       where the same convention for the finite-time integration step was used
       as above, i.e.,
        ${f}_S( \ \cdot \ \vert \ \cdot \ )$ refers to the r.h.s.\ of
        Eq.\ (\ref{eq:newMCTDH_S_diffequation}).
       This is the ``gauge correction'' step referred to above: As a result of
       this step, a new initial value ${\bm{S}}^{(\kappa)}(t_0)$ is generated,
       whose forward propagation according to
       Eq.\ (\ref{eq:newMCTDH_S_diffequation}) matches the
       $({\bm{\varphi}}^{(\kappa)} (t+\delta/2),
       {\bm{S}}^{(\kappa)}(t+\delta/2))$ combination
       obtained in the previous step. This
       ${\bm{S}}^{(\kappa)}(t_0)$ value defines the correct initial condition,
       in the new gauge, for the $A$ coefficient in the next step.
\item update of the $A$ vector at time $t_0$ according to Eq.\
         (\ref{eq:QRdecomposition_matrix}), {\it i.e.},
         $\boldsymbol{A}^{(\kappa)}(t_0) = \boldsymbol{Q}^{(\kappa)}(t_0)
         \boldsymbol{S}^{(\kappa)T}(t_0)$,
         where $\boldsymbol{S}^{(\kappa)T}(t_0)$ results from the preceding
         integration half-step
         while $\boldsymbol{Q}^{(\kappa)}$ was held fixed.
\end{itemize}
This sequence is iterated for each mode, with $\kappa = 1,2,  \dots f $.

\vspace*{0.2cm}

In the second step of the algorithm, the updated $A$ vector is integrated for a full
time step:
 \begin{equation*}
   \Delta {\bm{A}} = \int_{t_0}^{t_0+\delta} dt'\, {f}_A
   ({\bm{A}} (t') \ \vert \
   {\bm{\varphi}}^{(\kappa)}(t_0+\frac{\delta}{2}) )
\end{equation*}
where ${f}_A( \ \cdot \ \vert \ \cdot \ )$ refers to the r.h.s.\ of
Eq.\ (\ref{eq:regMCTDH_coeff_diffequation}).

\vspace*{0.2cm}

The third step of the algorithm is logically equivalent to the first one, with the difference that
the initial QR decomposition acts on the $A$ vector at time $t_0 + \delta$. For this reason,
the order of the integration of the SPFs and of ${\bm{S}}^{(\kappa)}$ is reversed:
\begin{itemize}[wide, labelwidth=!, labelindent=0pt]
 \item QR decomposition of the $A$ vector (Eq.\ (\ref{eq:QRdecomposition_matrix})) and construction of the
       corresponding mean--field potential (Eq.\ (\ref{eq:new_mean_field}))
       at time $t = t_0+ \delta$.
 \item backward integration of Eq.\ (\ref{eq:newMCTDH_S_diffequation}) for a half time--step
       towards a ($t_0+ \frac{1}{2} \delta$) value of ${\bm{S}}^{(\kappa)}$
       (``gauge correction''):
       \begin{equation*}
            \Delta {\bm{S}}^{(\kappa)}  =  \int_{t_0+\delta}^{t_0+\frac{\delta}{2}} dt'\, {f}_S
             ({\bm{S}}^{(\kappa)}(t') \ \vert \ \langle \tilde{\bm{H}}
             \rangle^{(\kappa)} (t_0+\delta)
                  {\bm{\varphi}}^{(\kappa)}(t_0+\frac{\delta}{2}) )
       \end{equation*}
       The updated ${\bm{S}}^{(\kappa)}$ is subsequently used to construct
       non--orthogonal SPFs $\tilde{\bm{\varphi}}^{(\kappa)}$ (\ref{eq:SPF_nonorthogonaldefinition}).
 \item forward integration of Eq.\ (\ref{eq:newMCTDH_spf_diffequation}) for a half time--step,
       to obtain new values of the non--orthogonal SPFs at time $t = (t_0 +
       \delta)$:
 \begin{equation*}
   \Delta \tilde{\bm{\varphi}}^{(\kappa)} = \int_{t_0+\delta/2}^{t_0+\delta} dt'\, {f}_{\varphi}
   (\tilde{\bm{\varphi}}^{(\kappa)} (t') \ \vert \ \langle \tilde{\bm{H}}
   \rangle^{(\kappa)} (t_0 + \delta) )
   \end{equation*}
 \item QR decomposition of the updated non--orthogonal SPFs (Eq.\
       (\ref{eq:SPF_nonorthogonaldefinition})) to obtain values of the orthonormal SPFs
       ${\bm{\varphi}}^{(\kappa)}$ and of the matrix ${\bm{S}}^{(\kappa)}$ at time
       $t = t_0 + \delta$, which is then used to update the $A$ vector (Eq.\
       (\ref{eq:QRdecomposition_matrix})).
\end{itemize}

Inspecting Fig. \ref{fig:integrationscheme}, it is evident that the first and second
half steps have a symmetric structure. In fact, one can be obtained from the other by inverting
the sequence of the operations.
The direction of time integration, however, needs to be maintained unaltered so that
both time steps globally result in a forward propagation of the SPFs.


\section*{References} 

\begin{thebibliography}{10}
\expandafter\ifx\csname url\endcsname\relax
  \def\url#1{\texttt{#1}}\fi
\expandafter\ifx\csname urlprefix\endcsname\relax\def\urlprefix{URL }\fi
\expandafter\ifx\csname href\endcsname\relax
  \def\href#1#2{#2} \def\path#1{#1}\fi

\bibitem{Meyer1990TheApproach}
H.-D. Meyer, U.~Manthe, L.~S. Cederbaum, {The Multi-Configurational
  Time-Dependent Hartree Approach}, Chem. Phys. Lett. 165 (1990) 73--78.

\bibitem{Manthe1992Wave-PacketNOCl}
U.~Manthe, H.-D. Meyer, L.~S. Cederbaum, {Wave-Packet Dynamics within the
  Multiconfiguration HartreeFramework: General Aspects and application to
  NOCl}, J. Chem. Phys. 97 (1992) 3199--3213.

\bibitem{Beck2000TheWavepackets}
M.~H. Beck, A.~J{\"{a}}ckle, G.~A. Worth, H.-D. Meyer, {The multiconfiguration
  time-dependent Hartree (MCTDH) method: a highly efficient algorithm for
  propagating wavepackets}, Phys. Rep. 324~(1) (2000) 1 -- 105.
\newblock \href {http://dx.doi.org/10.1016/S0370-1573(99)00047-2}
  {\path{doi:10.1016/S0370-1573(99)00047-2}}.

\bibitem{Wang2003MultilayerTheory}
H.~Wang, M.~Thoss, {Multilayer formulation of the
  multiconfigurationtime-dependent Hartree theory}, J. Chem. Phys. 119 (2003)
  1289--1299.

\bibitem{Manthe2008ASurfaces}
U.~Manthe, {A multilayer multiconfigurational time-dependent Hartree
  approachfor quantum dynamics on general potential energy surfaces}, J. Chem.
  Phys. 128 (2008) 164116.

\bibitem{Vendrell2011MultilayerPyrazine}
O.~Vendrell, H.-D. Meyer, {Multilayer multiconfiguration time-dependent Hartree
  method:Implementation and applications to a Henon-Heiles Hamiltonianand to
  pyrazine}, J. Chem. Phys. 134 (2011) 44135.

\bibitem{Koch2010DynamicalApproximation}
O.~Koch, C.~Lubich, {Dynamical Tensor Approximation}, SIAM J. Matrix Anal.
  Appl. 31~(5) (2010) 2360--2375.
\newblock \href {http://dx.doi.org/10.1137/09076578X}
  {\path{doi:10.1137/09076578X}}.

\bibitem{Lubich2013DynamicalTensors}
C.~Lubich, T.~Rohwedder, R.~Schneider, B.~Vandereycken, {Dynamical
  Approximation by Hierarchical Tucker and Tensor-Train Tensors}, SIAM J.
  Matrix Anal. Appl. 34~(2) (2013) 470--494.
\newblock \href {http://dx.doi.org/10.1137/120885723}
  {\path{doi:10.1137/120885723}}.

\bibitem{Bachmayr2016TensorEquations}
M.~Bachmayr, R.~Schneider, A.~Uschmajew, {Tensor Networks and Hierarchical
  Tensors for the Solution of High-Dimensional Partial Differential Equations},
  Found. Comput. Math. 16~(6) (2016) 1423--1472.
\newblock \href {http://dx.doi.org/10.1007/s10208-016-9317-9}
  {\path{doi:10.1007/s10208-016-9317-9}}.

\bibitem{Lubich2015TimeDynamics}
C.~Lubich, {Time Integration in the Multiconfiguration Time-Dependent Hartree
  Method of Molecular Quantum Dynamics}, Appl. Math. Res. Express 2015~(2)
  (2015) 311--328.

\bibitem{Kloss2017ImplementationApproach}
B.~Kloss, I.~Burghardt, C.~Lubich, {Implementation of a novel
  projector-splitting integrator for the multi-configurational time-dependent
  Hartree approach}, J. Chem. Phys. 146~(17) (2017) 174107.
\newblock \href {http://dx.doi.org/10.1063/1.4982065}
  {\path{doi:10.1063/1.4982065}}.

\bibitem{NoteQR}
{ Standard implementations of QR decomposition employ either the Modified
  Gram-Schmidt (MGS) algorithm or Householder transformations, in conjunction
  with column pivoting to handle the rank-deficient case \cite{Golub}.}

\bibitem{Golub}
G.~H. Golub, C.~F. {Van Loan}, {Matrix Computations}, John Hopkins University
  Press, 4th Ed., 2013.

\bibitem{HeidelbergMCTDH4}
H.-D. Meyer, F.~Gatti, G.~A. Worth (Eds.), {Multidimensional Quantum Dynamics:
  MCTDH Theory and Applications}, Wiley-VCH Verlag GmbH {\&} Co. KGaA,
  Weinheim, 2009.
\newblock \href {http://dx.doi.org/10.1002/9783527627400}
  {\path{doi:10.1002/9783527627400}}.

\bibitem{Meyer2012StudyingMethod}
H.-D. Meyer, {Studying molecular quantum dynamics with the multiconfiguration
  time-dependent Hartree method}, Wiley Interdiscip. Rev. Comput. Mol. Sci. 2
  (2012) 351.
\newblock \href {http://dx.doi.org/10.1002/wcms.87}
  {\path{doi:10.1002/wcms.87}}.

\bibitem{Manthe2015TheRevisited}
U.~Manthe, {The multi-configurational time-dependent Hartree approach
  revisited}, J. Chem. Phys. 142~(24) (2015) 244109.
\newblock \href {http://dx.doi.org/10.1063/1.4922889}
  {\path{doi:10.1063/1.4922889}}.

\bibitem{Conte2010AnDynamics}
D.~Conte, C.~Lubich, {An error analysis of the multi-configuration
  time-dependent Hartree method of quantum dynamics}, ESAIM: Math. Model.
  Numer. Anal. 44~(4) (2010) 759--780.
\newblock \href {http://dx.doi.org/10.1051/m2an/2010018}
  {\path{doi:10.1051/m2an/2010018}}.

\bibitem{Kato04}
T.~Kato, H.~Kono, {Time-dependent multiconfiguration theory for electronic
  dynamics of molecules in an intense laser field}, Chem. Phys. Lett. 533
  (2004) 392.

\bibitem{Zanghellini04}
J.~Zanghellini, M.~Kitzler, T.~Brabec, A.~Scrinzi, {Testing the
  multi-configuration time-dependent Hartree–Fock method}, J. Phys. B 37
  (2004) 763.

\bibitem{Nest05}
M.~Nest, T.~Klamroth, P.~Saalfrank, {The multiconfiguration time-dependent
  Hartree–Fock method for quantum chemical calculations}, J. Chem. Phys. 122
  (2005) 124102.

\bibitem{Hinz2016InstabilitiesHartree-Fock}
C.~M. Hinz, S.~Bauch, M.~Bonitz, {Instabilities and inaccuracies of
  multi-configuration time-dependent Hartree-Fock}, J. Phys. Conf. Ser. 696
  (2016) 012009.
\newblock \href {http://dx.doi.org/10.1088/1742-6596/696/1/012009}
  {\path{doi:10.1088/1742-6596/696/1/012009}}.

\bibitem{Burghardt1999ApproachesMethod}
I.~Burghardt, H.-D. Meyer, L.~S. Cederbaum, {Approaches to the approximate
  treatment of complex molecular systems by the multiconfiguration
  time-dependent Hartree method}, J. Chem. Phys.\href
  {http://dx.doi.org/10.1063/1.479574} {\path{doi:10.1063/1.479574}}.

\bibitem{Burghardt2003MulticonfigurationalBath}
I.~Burghardt, M.~Nest, G.~A. Worth, {Multiconfigurational system-bath dynamics
  using Gaussian wave packets: Energy relaxation and decoherence induced by a
  finite-dimensional bath}, J. Chem. Phys. 119~(11) (2003) 5364--5378.
\newblock \href {http://dx.doi.org/10.1063/1.1599275}
  {\path{doi:10.1063/1.1599275}}.

\bibitem{Burghardt2008MultimodePyrazine}
I.~Burghardt, K.~Giri, G.~A. Worth, {Multimode quantum dynamics using Gaussian
  wavepackets: The Gaussian-based multiconfiguration time-dependent Hartree
  (G-MCTDH) method applied to the absorption spectrum of pyrazine}, J. Chem.
  Phys. 129~(17) (2008) 174104.
\newblock \href {http://dx.doi.org/10.1063/1.2996349}
  {\path{doi:10.1063/1.2996349}}.

\bibitem{Romer2013Gaussian-basedTheory}
S.~R{\"{o}}mer, M.~Ruckenbauer, I.~Burghardt, {Gaussian-based
  multiconfiguration time-dependent Hartree: A two-layer approach. I. Theory},
  J. Chem. Phys. 138~(6) (2013) 064106.
\newblock \href {http://dx.doi.org/10.1063/1.4788830}
  {\path{doi:10.1063/1.4788830}}.

\bibitem{Worth2004AWavepackets}
G.~A. Worth, M.~A. Robb, I.~Burghardt, {A novel algorithm for non-adiabatic
  direct dynamics using variational Gaussian wavepackets}, Faraday Discuss.
  127~(0) (2004) 307.
\newblock \href {http://dx.doi.org/10.1039/b314253a}
  {\path{doi:10.1039/b314253a}}.

\bibitem{Worth2003FullWavepackets}
G.~A. Worth, I.~Burghardt, {Full quantum mechanical molecular dynamics using
  Gaussian wavepackets}, Chem. Phys. Lett. 368~(3-4) (2003) 502--508.
\newblock \href {http://dx.doi.org/10.1016/S0009-2614(02)01920-6}
  {\path{doi:10.1016/S0009-2614(02)01920-6}}.

\bibitem{Richings2015QuantumMethod}
G.~Richings, I.~Polyak, K.~Spinlove, G.~Worth, I.~Burghardt, B.~Lasorne,
  {Quantum dynamics simulations using Gaussian wavepackets: the vMCG method},
  Int. Rev. Phys. Chem. 34~(2) (2015) 269--308.
\newblock \href {http://dx.doi.org/10.1080/0144235X.2015.1051354}
  {\path{doi:10.1080/0144235X.2015.1051354}}.

\bibitem{Lubich2008FromAnalysis}
C.~Lubich, {From Quantum to Classical Molecular Dynamics: Reduced Models and
  Numerical Analysis}, European Mathematical Society Publishing House, Zuerich,
  Switzerland, 2008.
\newblock \href {http://dx.doi.org/10.4171/067} {\path{doi:10.4171/067}}.

\bibitem{Broeckhove1988OnPrinciples}
J.~Broeckhove, L.~Lathouwers, E.~Kesteloot, P.~Van~Leuven, {On the equivalence
  of time-dependent variational principles}, Chem. Phys. Lett. 149~(5-6) (1988)
  547--550.
\newblock \href {http://dx.doi.org/10.1016/0009-2614(88)80380-4}
  {\path{doi:10.1016/0009-2614(88)80380-4}}.

\bibitem{McLachlan1964AEquation}
A.~McLachlan, {A variational solution of the time-dependent Schrodinger
  equation}, Mol. Phys. 8~(1) (1964) 39--44.
\newblock \href {http://dx.doi.org/10.1080/00268976400100041}
  {\path{doi:10.1080/00268976400100041}}.

\bibitem{Kucar1987Time-dependentApproach}
J.~Kucar, H.-D. Meyer, L.~Cederbaum, {Time-dependent rotated hartree approach},
  Chem. Phys. Lett. 140~(5) (1987) 525--530.
\newblock \href {http://dx.doi.org/10.1016/0009-2614(87)80480-3}
  {\path{doi:10.1016/0009-2614(87)80480-3}}.

\bibitem{Raab2000OnPrinciple}
A.~Raab, {On the Dirac–Frenkel/McLachlan variational principle}, Chem. Phys.
  Lett. 319~(5-6) (2000) 674--678.
\newblock \href {http://dx.doi.org/10.1016/S0009-2614(00)00200-1}
  {\path{doi:10.1016/S0009-2614(00)00200-1}}.

\bibitem{Lubich2004OnDynamics}
C.~Lubich, {On variational approximations in quantum molecular dynamics}, Math.
  Comput. 74~(250) (2004) 765--780.
\newblock \href {http://dx.doi.org/10.1090/S0025-5718-04-01685-0}
  {\path{doi:10.1090/S0025-5718-04-01685-0}}.

\bibitem{MeyerWang}
H.~Meyer, H.~Wang, {On regularizing the MCTDH equations of motion}, J. Chem.
  Phys. 148~(5) (2018) 124105.

\bibitem{Tuckerman2010StatisticalSimulations}
M.~Tuckerman, {Statistical mechanics and molecular simulations}, Oxford
  Graduate Texts, Oxford University Press, Oxford, 2010.

\bibitem{Feit1982SolutionMethod}
M.~Feit, J.~Fleck, A.~Steiger, {Solution of the Schr{\"{o}}dinger equation by a
  spectral method}, J. Comput. Phys. 47~(3) (1982) 412--433.
\newblock \href {http://dx.doi.org/10.1016/0021-9991(82)90091-2}
  {\path{doi:10.1016/0021-9991(82)90091-2}}.

\bibitem{Beck1997}
M.~H. Beck, H.-D. Meyer, {An efficient and robust integration scheme for the
  equations of motion of the multiconfiguration time-dependent Hartree (MCTDH)
  method}, Z. Phys. 42~(1) (1997) 113--129.

\bibitem{Haegeman16}
J.~Haegeman, C.~Lubich, I.~Oseledets, B.~Vandereycken, F.~Verstraete, {Unifying
  time evolution and optimization with matrix product states}, Phys. Rev. B 94
  (2016) 165116.

\bibitem{Haegeman13}
J.~Haegeman, T.~J. Osborne, F.~Verstraete, {Post-matrix product methods: To
  tangent space and beyond}, Phys. Rev. B 88 (2013) 075133.

\bibitem{Haegeman11}
J.~Haegeman, J.~I. Cirac, T.~J. Osborne, I.~Pizorn, H.~Verschelde,
  F.~Verstraete, {Time-dependent variational principle for quantum lattice},
  Phys. Rev. Lett. 107 (2011) 070601.

\end{thebibliography}

\end{document}